\documentclass[10pt]{iopart}
\usepackage{iopams}
\usepackage{graphicx}
\usepackage{cite}
\usepackage{color}
\usepackage{multirow}
\usepackage[section]{placeins}

\usepackage[deletedmarkup=xout]{changes}

\definechangesauthor[color=red]{ZD}

\definechangesauthor[color=blue]{SD}

\begin{document}

\setlength{\parindent}{0pt}

\title[Equilibration of electron swarms]{Experimental observation and simulation of the equilibration of electron swarms in a scanning drift tube\\
}

\author{Z. Donk\'o$^{1}$, P. Hartmann$^{1}$, I. Korolov$^{2}$, V. Jeges$^{1}$, D. Bo\v{s}njakovi\'c$^{3}$, S. Dujko$^{3}$}

\address{$^1$Institute for Solid State Physics and Optics, Wigner Research Centre for Physics, Hungarian Academy of Sciences, 1121 Budapest, Konkoly Thege Mikl\'os str. 29-33, Hungary\\
}
\address{$^2$Department of Electrical Engineering and Information Science, Ruhr-University Bochum, D-44780,
Bochum, Germany}

\address{$^3$Institute of Physics, University of Belgrade, Pregrevica 118, 11080 Belgrade, Serbia}

\ead{donko.zoltan@wigner.mta.hu}

\begin{abstract}

We investigate the spatially and temporally resolved electron kinetics in a homogeneous electric field in argon gas, in the vicinity of an emitting boundary. This (transient) region, where the electron swarm exhibits non-equilibrium character with energy gain and loss processes taking place at separate positions (in space and time), is monitored experimentally in a scanning drift tube apparatus. Depending on the strength of the reduced electric field we observe the equilibration of the swarm over different length scales, beyond which the energy gain and loss mechanism becomes locally balanced and transport properties become spatially invariant. The evolution of the electron swarm in the experimental apparatus is also described by Monte Carlo simulations, of which the results are in good agreement with the experimental observations, over the domains of the reduced electric field and the gas pressure covered.

\end{abstract}


\maketitle

\section{Introduction}

The description of charged particle transport in plasma modelling is often based on {\it transport coefficients} (e.g. mobility and diffusion coefficients) that are functions of the reduced electric field (electric field to gas density ratio, $E/N$). These coefficients can be determined experimentally in {\it swarm experiments} in which a cloud of charged particles (e.g. electrons) moves under the influence of a homogeneous electric field. The basic tools for measurements of these coefficients have been {\it drift tubes}, e.g., \cite{DT1,DT2,DT3,DT4,DT5}, which can operate in different modes (steady-state or pulsed) and give various transport coefficients \cite{swarm1}. Obtaining precise transport coefficients experimentally also aids the optimisation of cross section sets \cite{cross1,cross2,cross3}. 

Swarm experiments aimed at the determination of transport coefficients have to be conducted under the conditions of {\it equilibrium transport}, where the effects of boundaries are negligible and gradients are weak and the electron velocity distribution function (VDF, $f({\bf v})$) is uniquely defined by $E/N$. As the VDF of the "initial" electrons (created, e.g., by ultraviolet radiation) in any experimental system is different from the equilibrium VDF, the swarm needs a certain length to equilibrate, during which length the energy (momentum) gain and loss mechanisms get balanced (see, e.g., \cite{equi,Donko2011}). This equilibration domain (within which the transport has "non-hydrodynamic" or "non-local" character) should ideally be excluded from the region from which data for the determination of transport coefficients is collected as here the characteristics of the swarm vary spatially despite the fact the electric field that drives the transport, is homogeneous. Excluding the equilibration region from the measurements is, however, normally not possible, but one has to ensure that the effects of swarm equilibration in the drift region are minimal, e.g., by setting the drift length significantly longer than the equilibration length. Simulations of the electron transport are indispensable tools for checking this condition.

The {\it equilibration length} of any swarm depends on the type of its constituents (electrons / ions), the electric field, the gas pressure, and the types of collision processes between the charged particles and the atoms/molecules of the buffer gas. In the following we focus only on electron swarms. In the case of atomic gases, at very low $E/N$ values, where the electron energy cannot reach the threshold for inelastic processes (which is typically several eV), only elastic collisions take place. At very high $E/N$ values, where several inelastic channels are open, the electrons can lose several discrete values of energy in various excitation events, and, in ionisation processes their energy loss can vary continuously. Under these conditions the equilibration of the swarms proceeds quickly, over a short spatial domain. There exist, however, a "window" of $E/N$ values, typically in the range of several tens of Td-s (1 Td = 10$^{-21}$ V\,m$^2$), where the equilibration takes place over an extended spatial scale \cite{per1,per2,per3,per4}. The reason for this is that the electrons gain energy slowly (due to the relatively low electric field) and predominantly excite only the lowest excited state(s). In these conditions, the energy gain - energy loss cycle may repeat many times, the local swarm characteristics exhibit a periodic spatial dependence, before stationary state forms. In the case of molecular gases, the equilibration of the electron swarms proceeds more quickly due to the existence of various types of excitations processes (rotational, vibrational and electronic excitation), some of them having low threshold energies \cite{LiWR2002,DujkoWPR2011}. It should be noted that additional control of the spatial relaxation of electrons under the steady-state conditions can be achieved using a magnetic field \cite{LiRW2006,DujkoWPR2011}.

The aim of this work is to examine the equilibration of electron swarms experimentally, in a drift tube apparatus that allows the observation of the spatio-temporal development of the particle could \cite{co2}. We do not target here the determination of transport coefficients. Parallel to the experimental studies we also carry out simulations at the particle level, to illustrate the phenomenon of swarm equilibration and to describe particle motion in the actual experimental system. Our studies are conducted using argon as a buffer gas.

In section \ref{sec:methods} we give a brief description of the experimental system and outline the basics of the Monte Carlo simulation method that we use as a computational tool for our studies of swarm equilibration. In order to illustrate the phenomenon of swarm equilibration, in general, we first present a set of simulation results for a simple setting with a plane-parallel electrode configuration, for steady-state and time-dependent conditions, in section \ref{sec:results1}. Subsequently, in section \ref{sec:results2}, we turn to the presentation of experimental results and we compare these results with those obtained from simulations of the experimental system. Subsequently, in section \ref{sec:detector}, we also present additional simulation results that aid the understanding the operation of the detector of the drift tube. Section \ref{sec:sum} summarises our findings. 

\section{Methods}

\label{sec:methods}

We investigate the equilibration of electron swarms both experimentally, in a scanning drift tube apparatus \cite{rsi} and via particle level simulations based on the Monte Carlo technique. The latter provides a description of particle transport at the level of kinetic theory, thus it is expected to account fully for the behaviour of the swarms under the specific (usually non-hydrodynamic) conditions considered here. 

Full description of the experimental apparatus has been given in Ref. \cite{rsi}, thus only the main features of the setup are presented below, in section \ref{sec:exp}. The basics of the simulation method are outlined in section \ref{sec:sim}.

\subsection{Experimental system}

\label{sec:exp}

\begin{figure}[ht ]
\begin{center}
\includegraphics[width =0.8\textwidth]{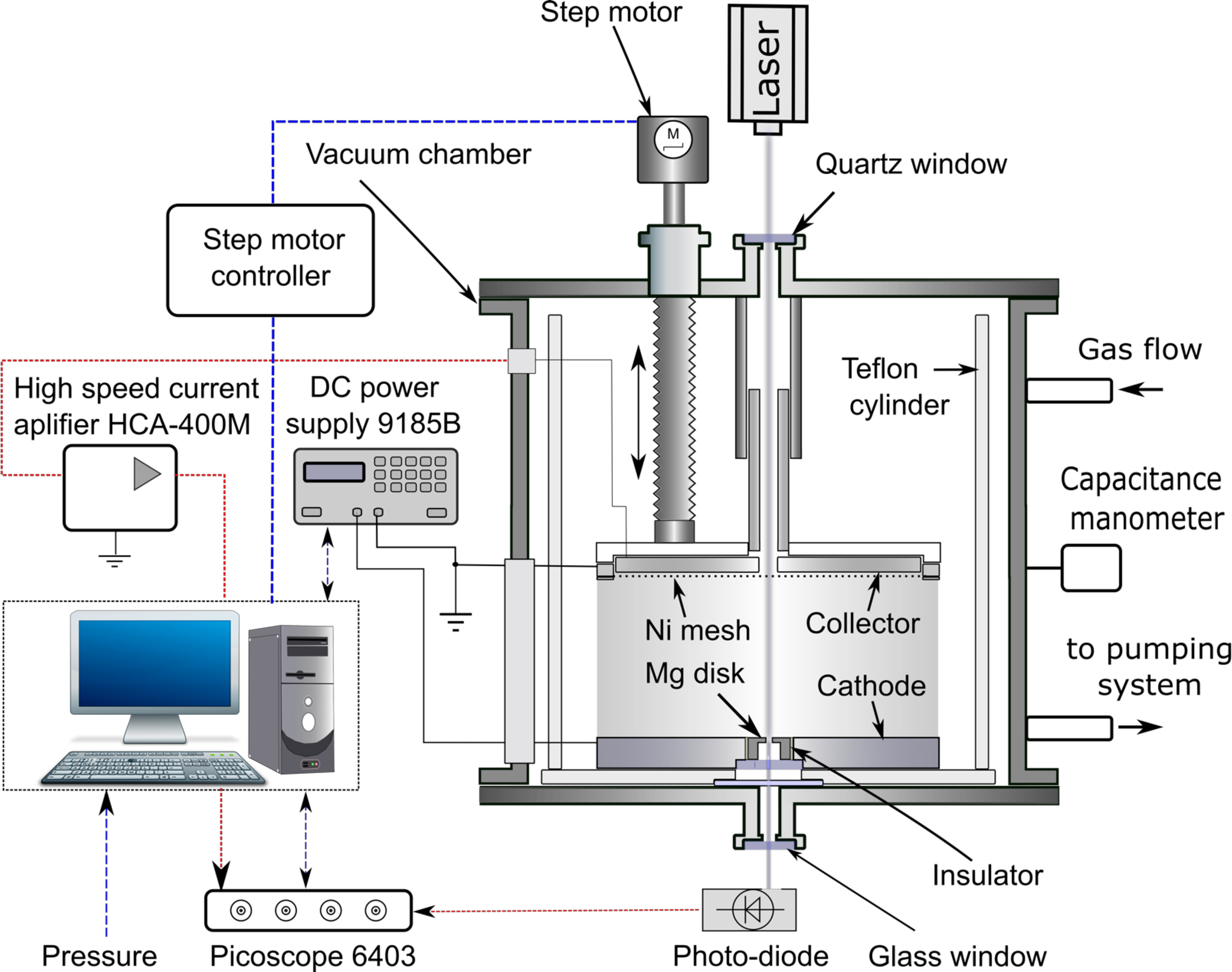}
\caption{Simplified scheme of the experimental setup.}
\label{fig:expsetup}
\end{center}
\end{figure}

The simplified scheme of the experimental setup is shown in figure \ref{fig:expsetup}. The drift tube is situated within a stainless steel vacuum chamber that is evacuated by a turbomolecular pump backed with a rotary pump, down to a level of $\sim$ 10$^{-7}$ mbar. A feedthrough with a quartz window allows the 1.7 $\mu$J-energy, 5 ns-long pulses of a frequency-quadrupled diode-pumped YAG laser to fall on the surface of a Mg disk used as photoemitter. This disk is mounted at the center of a stainless steel electrode (having a diameter of 105 mm) serving as the cathode of the drift tube, which is connected to a BK Precision 9185B power supply to establish the accelerating voltage for the swarm that moves towards the detector, situated at a distance $L_1$ from the emitter. The detector consists of a grounded nickel mesh (with $T$ = 88\% "geometric" transmission and 45 lines/inch density) and a stainless steel collector electrode that is situated at 1 mm distance behind the mesh. The grid and the collector are moved together by a step motor connected to a micrometer screw mounted via a vacuum feedthrough to the vacuum chamber. The distance between the cathode and the mesh can be set within a range of $L_1$ = 7.8 -- 58.3 mm. The electric field is kept constant during the scanning process by automatically adjusting the cathode-mesh voltage, according to their actual distance. In the experiments presented here, we used 53 equidistant positions within the accessible range of $L_1$ given above.

The data collection is triggered with a photodiode, using a part of the laser light that passes through a hole in the magnesium disk and leaves the chamber via a window mounted on its bottom. The current of the detector system is generated by the moving charges within the mesh-collector gap (see below). This current is amplified by a high speed current amplifier (type Femto HCA-400M) connected to the collector, with a virtually grounded input and is recorded by a digital oscilloscope (type Picoscope 6403B) with sub-ns time resolution. During the experiments a slow ($\sim$sccm) flow of (6.0 purity) argon gas is established by a flow controller, the gas pressure inside the chamber is measured by a Pfeiffer CMR 362 capacitive gauge. The experiment is fully controlled by a computer using LabView software. 

\begin{figure}[ht ]
\begin{center}
\includegraphics[width =0.55\textwidth]{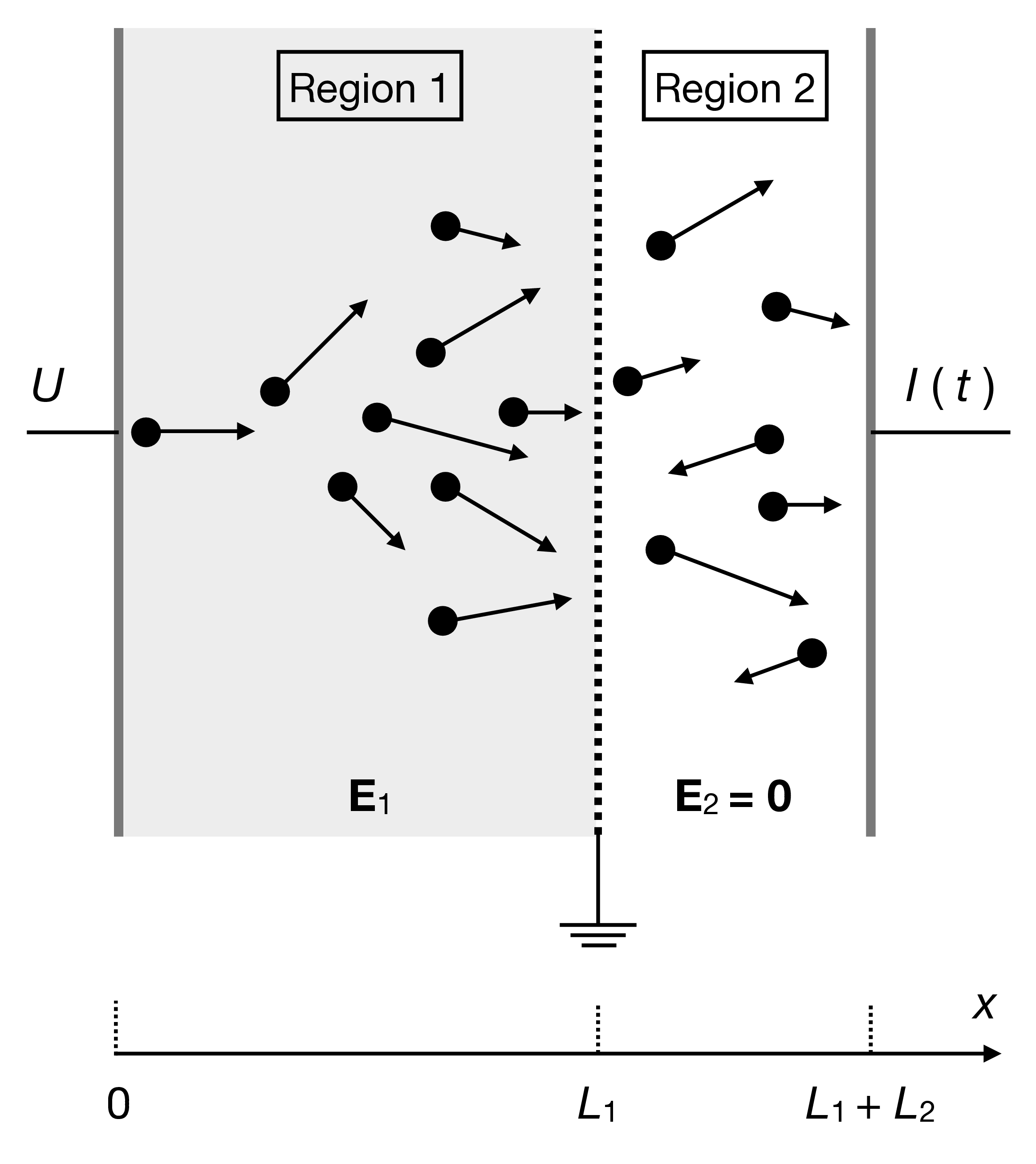}
\caption{Scheme of the two regions in the drift tube. A negative high voltage $U$ is applied at the cathode. In the experimental system $L_1$ can be changed between 7.8 mm and 58.3 mm, the distance $L_2$ is fixed at 1 mm. As the current $I(t)$ is measured with an amplifier that has a virtually grounded input, Region 2 is field-free. The measured current is generated by the moving electrons within this region.}
\label{fig:simple-setup}
\end{center}
\end{figure}

The operation of the drift tube described above can be understood by the simplified configuration shown in figure~\ref{fig:simple-setup}. The system consists of two regions, in Region 1 (shaded part in figure~\ref{fig:simple-setup}) a homogeneous electric field ${\bf E}_1= U/L_1$ (where $U$ is the voltage applied to the cathode (emitter)) accelerates the swarm of electrons, initiated at $x=0$, by the short-pulse laser. As both the mesh and the collector reside at ground potential, the electric field in Region 2, ${\bf E}_2$, is zero. The measured current at the collector is generated by the moving electrons within Region 2 \cite{SH} that create a displacement current, i.e., the measured current at a given time $t$ is 
\begin{equation}
I(t) = c \sum_{k} v_{x,k}(t),
\label{eq:ramo}
\end{equation}
where $c$ is a constant, the summation goes over all the electrons being present in Region 2 at time $t$, and $v_{x,k}$ is the velocity component of the $k$-th electron in the $x$ direction. We note that the motion of the electrons within Region 1 has no influence on this current.

\subsection{Simulation method}

\label{sec:sim}

We use simulations of electron swarms for three different purposes:
\begin{itemize}
\item{{\it to illustrate the general features of swarm equilibration within a plane-parallel electrode gap:} in these simulations electrons are emitted from the negatively biased cathode at $x$ = 0 and their tracing continues until the electrons are absorbed at the grounded anode at $x=L$, that has a reflectance $R$ for the electrons. We investigate both steady-state and time-dependent cases. In the steady-state case we illustrate the behaviour of the swarm by presenting the average velocity and the mean electron energy as a function of position. In the time-dependent case we show the spatio-temporal evolution of the density of the swarm, for different $E/N$ values.}
\item{{\it to describe the experimental system}: in this case we adopt the model geometry shown in figure~\ref{fig:simple-setup}. An electric field is applied only in Region 1, between the cathode and the mesh. The electrons can pass through the mesh with a probability that equals its geometric transmission ($T$ = 88\%). Electrons interacting with the mesh can be absorbed / elastically reflected with given probabilities. Electrons reaching the collector can as well be absorbed / elastically reflected with given probabilities.}
\item{{\it to study the sensitivity of the detector as a function of electron energy and gas pressure}: in this study we inject electrons with given energies into Region 2 (the "detector gap", see figure~\ref{fig:simple-setup}) and analyse the response of the detector as a function of these parameters.}
\end{itemize} 

Our simulations are based on the conventional Monte Carlo approach. Electrons are emitted from the cathode (situated at $x$ = 0) at $t=0$, with an initial energy of 1 eV. The typical number of initial electrons is in the order of 10$^5$--10$^6$. The electrons move under the influence of a homogeneous electric field, or in a field-free region, while interacting with the background gas via collision processes: elastic and inelastic (excitation and ionisation) collisions. Between collisions the electrons move on trajectories defined by their equations of motion that are discretised and solved with a time step $\Delta t$:
\begin{eqnarray}
x(t+\Delta t) = x(t) + v_x(t) \Delta t + \frac{1}{2} a \Delta t^2, \\
v_x(t+\Delta t) = v_x(t) + a \Delta t,
\end{eqnarray}
with $a= -\frac{eE}{m}$, where $e$ is the elementary charge and $m$ is the electron mass. The directions ($y$ and $z$) perpendicular to the direction of the electric field are not resolved.

The probability of a collision to take place after $\Delta t$ is given as
\begin{equation}
P(\Delta t)=1-\exp[-N {\sigma_{\rm T}(v)v}\Delta t],
\end{equation}
where $N$ is the gas density, $\sigma_{\rm T}$ is the total scattering cross section, and $v$ is the velocity of the electron (i.e., we use the cold-gas approximation, where target atoms are at rest). The simulation time step is in the order of 10$^{-12}$ s.

Comparison of $P(\Delta t)$ with a random number $r_{01}$ (having a uniform distribution over the [0,1) interval) allows deciding about the occurrence of a collision: if $r_{01} \leq P(\Delta t)$ a collision is simulated. The type of collision is determined in a random manner. The probability of a process $s$ at a given energy $\varepsilon$ is given by:
\begin{equation}
P_s = \frac {\sigma_s(\varepsilon)} {\sigma_{\rm T}(\varepsilon)},
\end{equation} 
where $\sigma_s(\varepsilon)$ is the cross section of the $s$-th process. In our simulations the cross sections are adopted  from \cite{Hayashi}. Collisions are assumed to result in isotropic scattering. Accordingly, we use the elastic momentum transfer cross section. For a given gas pressure, the background gas number density is calculated assuming the temperature of 300 K.

\section{Results}

\subsection{Swarm equilibration under steady-state and time-dependent conditions}

\label{sec:results1}

The relaxation of electron swarms is first illustrated for steady-state systems (termed as "Steady State Townsend" (SST) scenario \cite{per2,Robson,SakaiTS1977,StojanovicP1998} in swarm physics). In these simulations we assume a simple plane-parallel electrode configuration and consider a continuous source of electrons at the cathode. The electrode gap is chosen to be the largest distance of the cathode and the mesh in the experiment, $L$ = 58.3 mm. 

\begin{figure}[h!]
\begin{center}
\footnotesize{(a)}\includegraphics[width=0.45\textwidth]{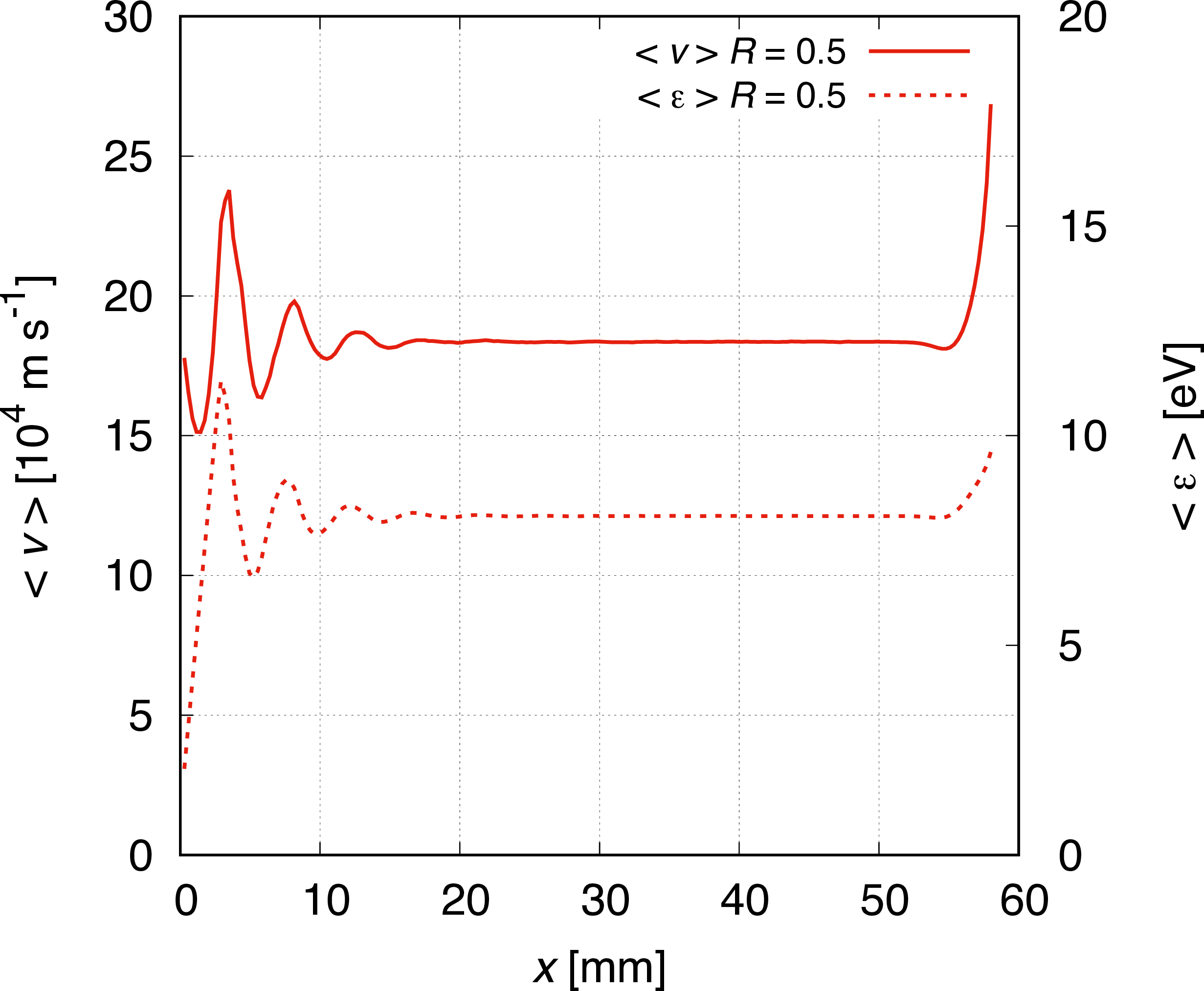}~~~
\footnotesize{(b)}\includegraphics[width=0.45\textwidth]{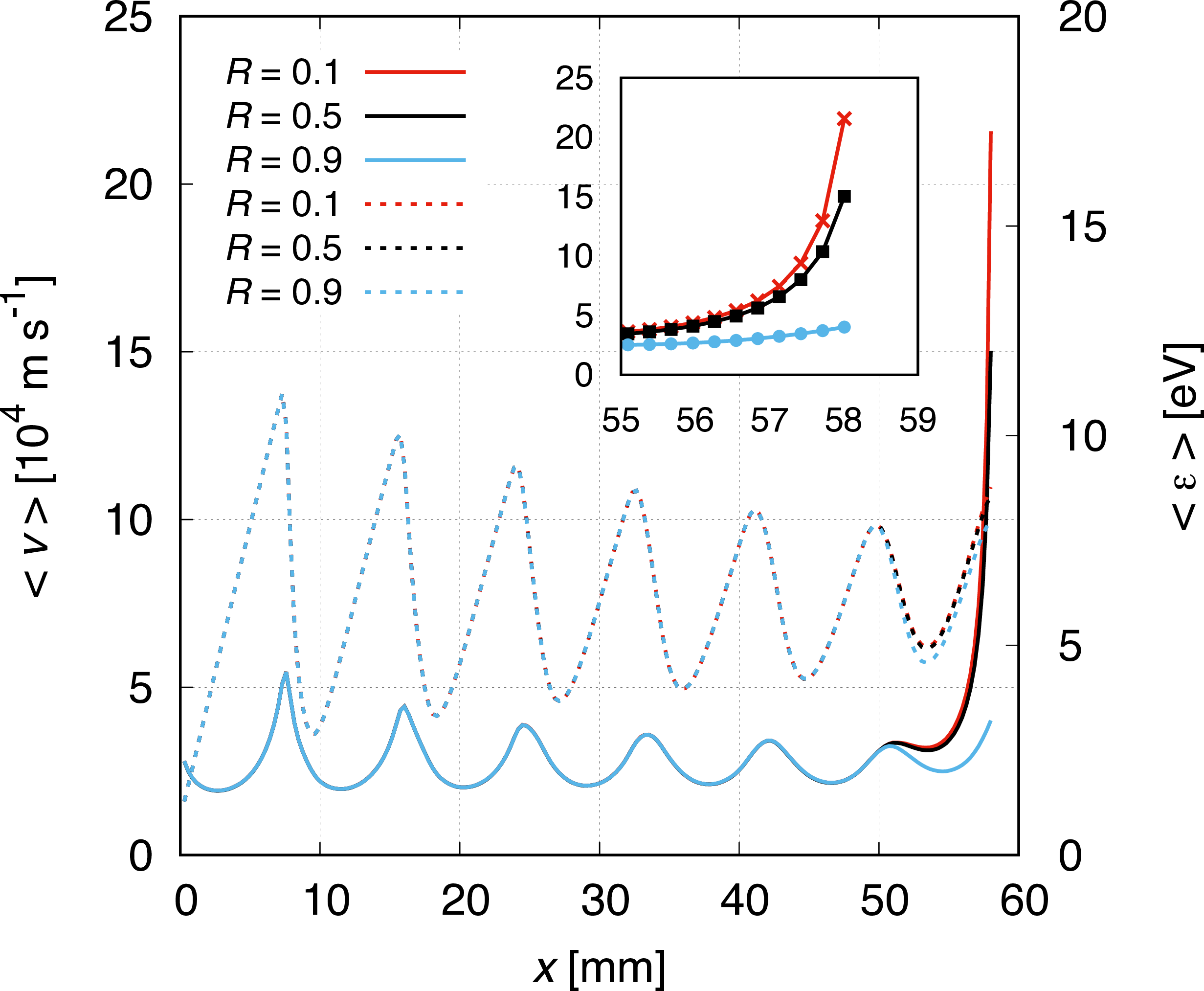}~~~
\caption{Average velocity (solid lines, left scale) and mean energy (dashed lines, right scale) of the electrons at (a) 300 Td ($p$ = 50 Pa) and (b) 30 Td ($p$ = 200 Pa), in the steady-state case. The cathode is situated at $x$ = 0 mm, while the anode is at $L$ = 58.3 mm. $R$ denotes the electron reflection coefficient of the anode. The inset in (b) shows the effect of $R$ on the average velocity in the near-anode region.}
\label{fig:clear1}
\end{center}
\end{figure}

Figure \ref{fig:clear1} shows the average velocity and the mean energy of the electrons as a function of position in the electrode gap, for steady-state conditions. Panel (a) shows the results for $E/N$ = 300 Td. The equilibrium transport, with transport properties specific to a given $E/N$, is established beyond a certain distance. Within the "transient region" the local transport coefficients (like $\langle v \rangle$ and $\langle \varepsilon \rangle$) and the velocity distribution function, $f({\bf v})$, change with position.  Here, the swarm relaxes over a length of $\approx$\,20 mm, beyond this distance from the cathode the transport acquires equilibrium character, the transport parameters reach constant values and $f({\bf v})$ takes a steady shape (while its magnitude grows according to the increase of the electron density, due to ionising collisions). These characteristics become, however, perturbed again near the anode that is normally partially reflecting/absorbing for the electrons. The data shown here were obtained with a reflection coefficient of $R$ = 0.5 (that we assume to be independent of the electron energy and angle of incidence at the surface). As part of the electrons is absorbed by the anode, in its vicinity the $f({\bf v})$ distribution function is depleted in the $v_x < 0$ domain. This results in a significant increase of the average velocity and a moderate increase of the mean electron energy within a distance of a few mm-s from the anode. 

At a lower $E/N$ value of 30 Td, the relaxation of the swarm requires a notably longer distance, as indicated in figure \ref{fig:clear1}(b). In this case even the full length, $L$, is too short for the swarm to acquire the equilibrium character, $\langle v \rangle$ and $\langle \varepsilon \rangle$ exhibit oscillations over the whole electrode gap. For this $E/N$ value, simulations were carried out with different reflection coefficients. As expected, the mean velocity becomes more perturbed (increased) near the anode when $R$ is decreased. As in the experiments the measured current  originates from the motion of the electrons in the $x$-direction near the collector (in Region 2), the above observations have consequences on the performance of the experimental system. 

\begin{figure}[h!]
\begin{center}
\footnotesize{(a)}\includegraphics[width=0.45\textwidth]{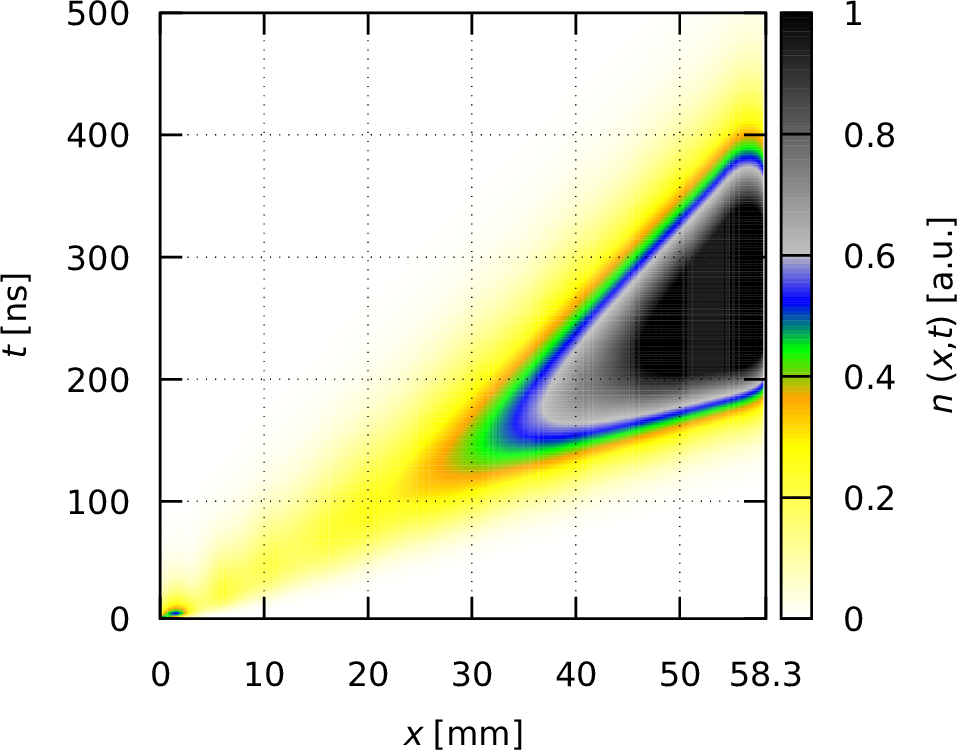}~~~
\footnotesize{(b)}\includegraphics[width=0.45\textwidth]{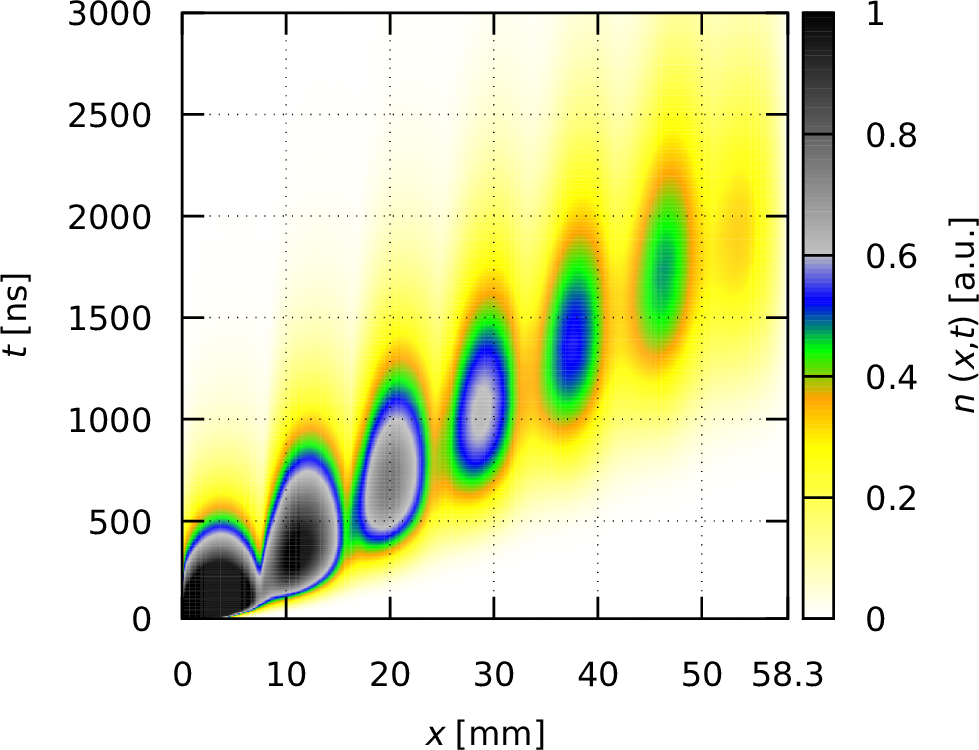}~~~
\caption{The evolution of the electron density in space and time for time-dependent conditions (swarm initiated at $x$ = 0 mm and $t$ = 0 ns) at (a) 300 Td ($p$ = 50 Pa) and (b) 30 Td ($p$ = 200 Pa). The cathode is situated at $x$ = 0 mm, while the anode is at $L$ = 58.3 mm.}
\label{fig:tof1}
\end{center}
\end{figure}

Next, we turn to time-dependent conditions: figure \ref{fig:tof1} shows the spatio-temporal evolution of the electron density for two different values of $E/N$, same as above. In this case electrons are emitted in the MC simulations from the cathode (situated at $x=0$) at time $t=0$.  Panel (a) displays the case of 300 Td ($p$ = 50 Pa). For this $E/N$ we observe a smooth development of the (density of the) particle cloud. Three basic effects are visible in this plot: (i) the centre of mass of the cloud {\it drifts} to higher $x$ values with increasing time, (ii) with increasing time we observe an increasing width of the cloud due to {\it diffusion}, and (iii) the density increases with position as a consequence of {\it ionising collisions}. Except from the vicinity of the cathode no structures can be seen in the density distribution, unlike in the case of 30 Td ($p$ = 200 Pa), shown in panel (b) of figure \ref{fig:tof1}. Here, similar to the steady-state case, a significant spatial variation of the swarm evolution is found. The "lobes" in figure \ref{fig:tof1}(b) represent local density peaks, where electrons accumulate. These localised maxima in space and time are created as a consequence of the repeating energy gain - energy loss cycles. For the given conditions the voltage over the gap is 84.5 V, that gives an electric field of 1.45 V\,mm$^{-1}$. For the distance of the peaks, $\Delta x \approx$ 9 mm, a potential drop of $\approx$ 13 V over the length scale of $\Delta x$ is obtained, which corresponds closely to lowest excitation energies of argon atoms. As the electrons can excite a number of energy levels with different threshold energies, their energy gain/loss cycles are not completely synchronised and therefore the density modulation decreases while the swarm moves, and after a certain distance the modulation disappears, and the swarm takes the equilibrium character. 

\subsection{Swarm equilibration in the drift tube - experiment vs. simulation}

\label{sec:results2}

Following the brief introduction to the equilibration phenomenon, now we turn to the presentation of experimental results confirming this behaviour by direct measurements on electron swarms in argon, and to the comparison of the experimental results with simulation data obtained at identical conditions. This comparison is carried out in terms of the measured / computed currents. 

\begin{figure}[h!]
\footnotesize{(a)}\includegraphics[width=0.45\textwidth]{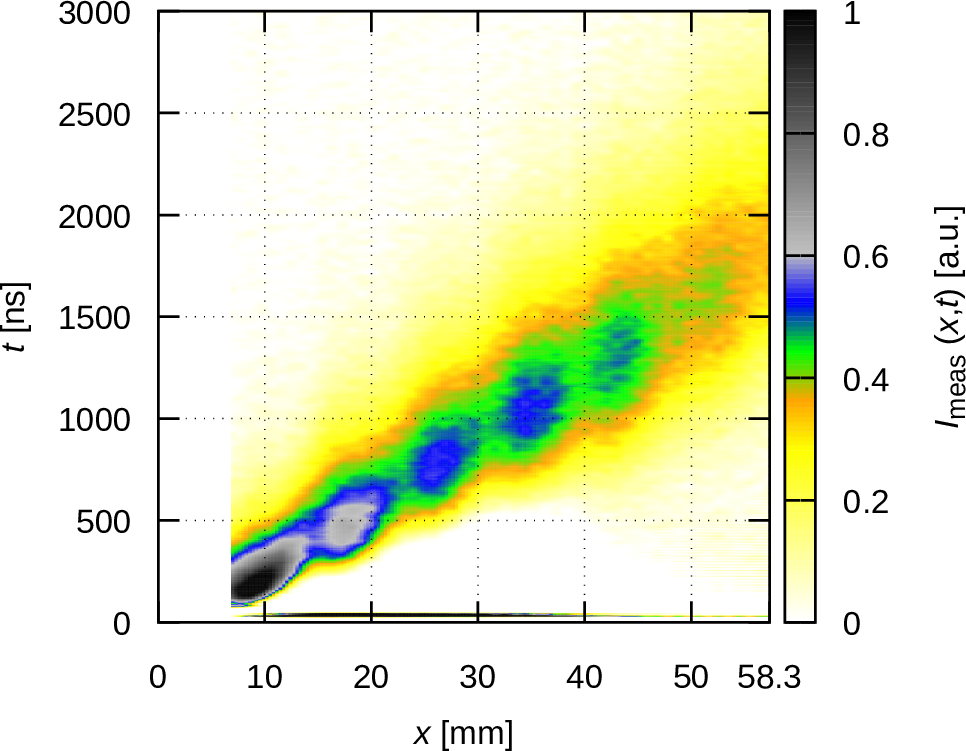}~
\footnotesize{(d)}\includegraphics[width=0.45\textwidth]{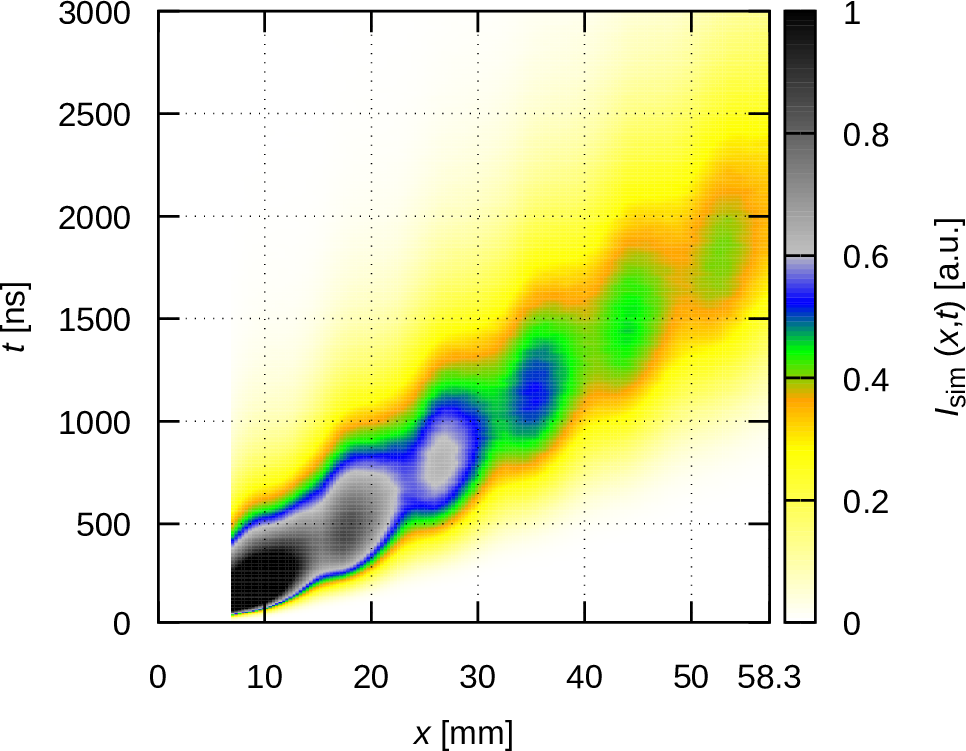}\\
\footnotesize{(b)}\includegraphics[width=0.45\textwidth]{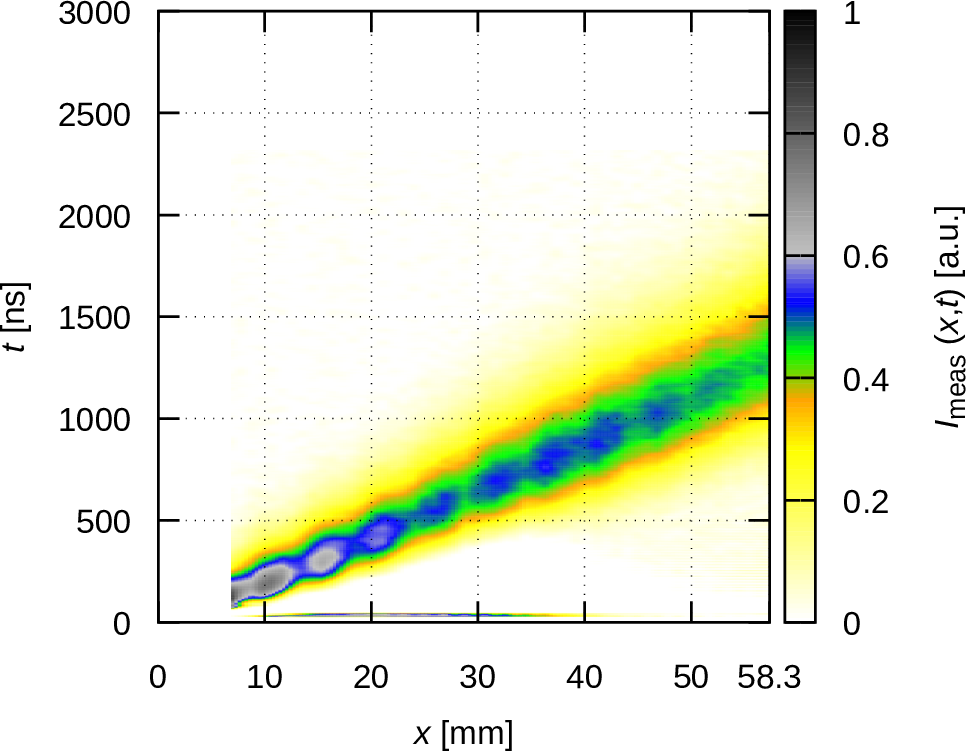}~
\footnotesize{(e)}\includegraphics[width=0.45\textwidth]{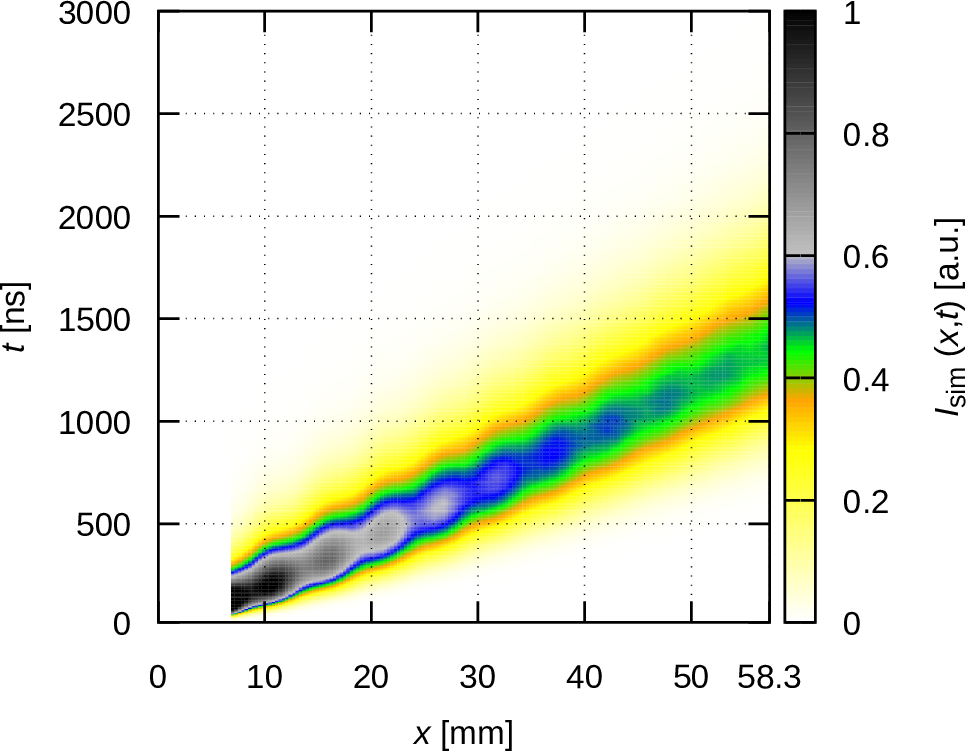}\\
\footnotesize{(c)}\includegraphics[width=0.45\textwidth]{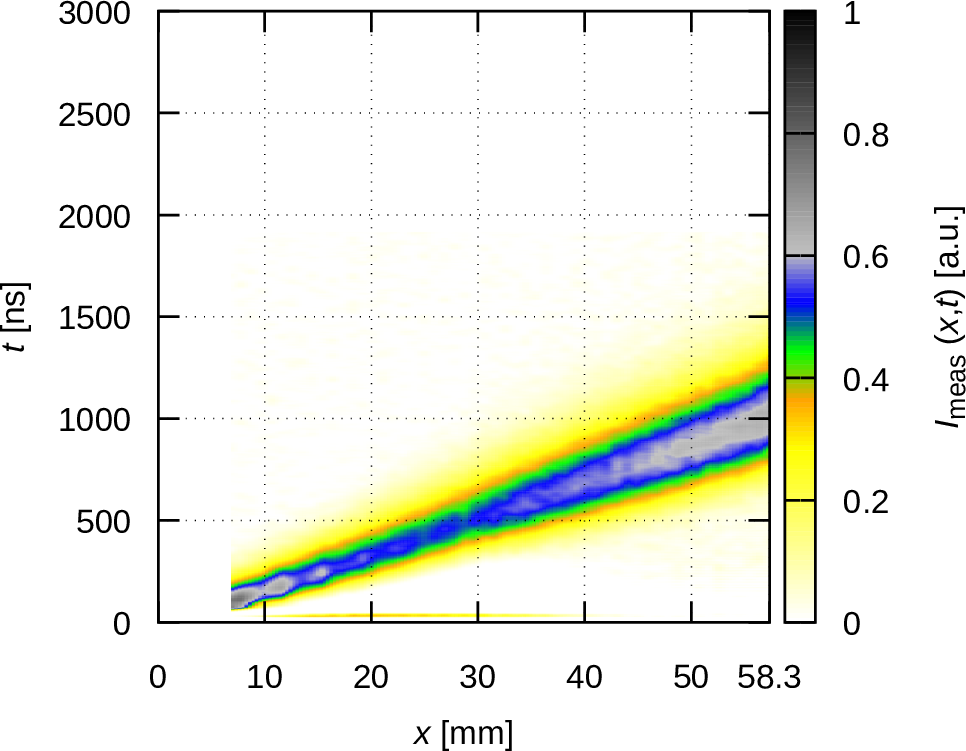}~
\footnotesize{(f)}\includegraphics[width=0.45\textwidth]{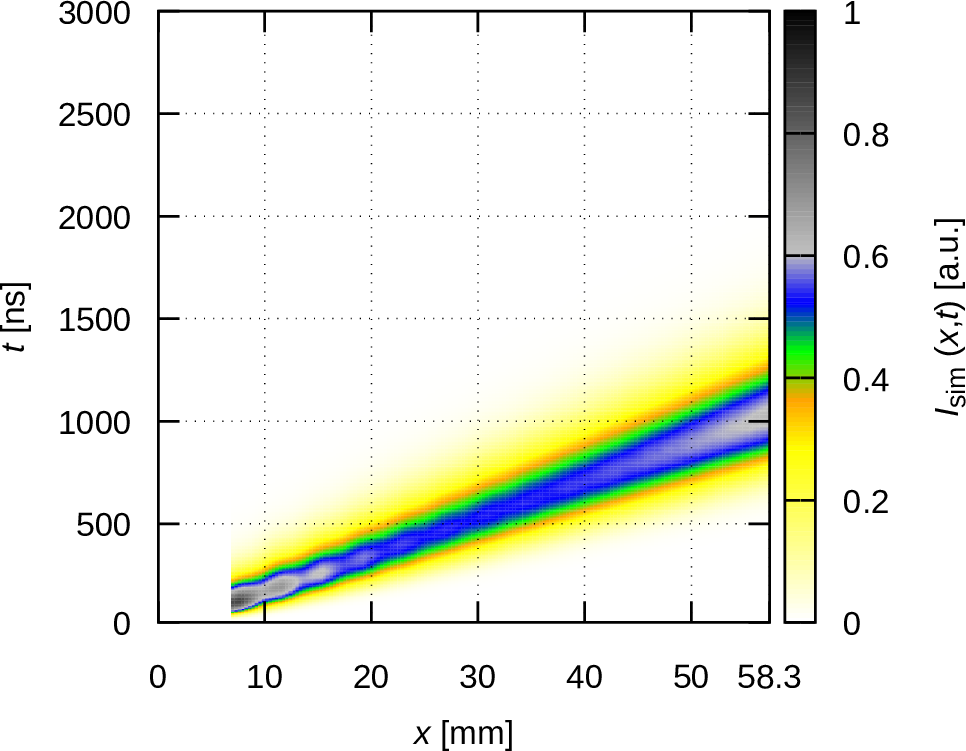}\\
\caption{Experimentally recorded detector current (a-c), at $E/N$ values of 30 Td, 50 Td and 70 Td, respectively, and (d-f) corresponding MC simulation results. $p$ = 200 Pa for all results.}
\label{fig:res1}
\end{figure}

We start with the presentation of the experimentally recorded "swarm maps" and the corresponding simulation results. Figure \ref{fig:res1} displays the experimental data in the left column (panels (a) (b), and (c)) obtained at 30 Td, 50 Td, and 70 Td values of the reduced electric field, respectively, at a fixed Ar pressure of $p$ = 200 Pa. These swarm maps have been generated by measuring the $I(t)$ current of the detector at 53 equidistant values within the $L_1$ = 7.8 -- 58.3 mm range of drift distances, and by merging these sets of data.

At the lower $E/N$ values the maps clearly show sequences of "lobes" that correspond to maxima of the measured currents, localised in both space and time. Taking the $E/N$ = 30 Td case as an example, the distance of the lobes in space is again approximately $\Delta x$  = 9 mm, as in the case of the theoretical results for the swarm density, shown in figure \ref{fig:tof1}(b). Note, however, that while in the density distribution maxima occur e.g., at about 29 mm and 38 mm, the measured current peaks at approximately 27 mm and 35 mm, i.e. the peaks are shifted by about 2 mm. The reason for this shift will be discussed later, based on an analysis of the electron trajectories in the detector region. When, however, the experimentally obtained map (of the detector current) is compared with that obtained from the simulation of the experimental configuration, a very good agreement is obtained both in terms of the structure of the map as well as in the precise positions of the maxima. This confirms the validity of the model and the correct description of the system by the simulation.

With increasing $E/N$, the distance of the lobes decreases as dictated by the above condition (at a fixed pressure). At $E/N$ = 50 Td a clear sequence of lobes can still be resolved (figure \ref{fig:res1}(b) and (e)), while at 70 Td signatures of periodic structures can still be seen within the first half of the drift distance, while the second half of the gap shows a smooth distribution (figure \ref{fig:res1}(c) and (f)).

\begin{figure}[h!]
\footnotesize{(a)}\includegraphics[width=0.45\textwidth]{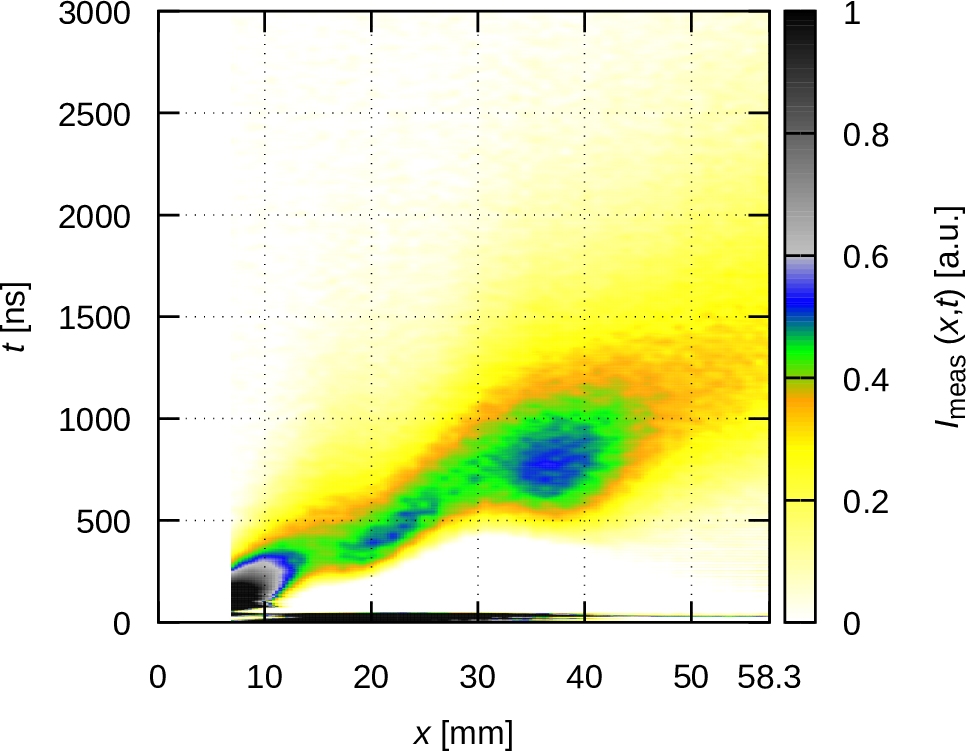}~
\footnotesize{(c)}\includegraphics[width=0.45\textwidth]{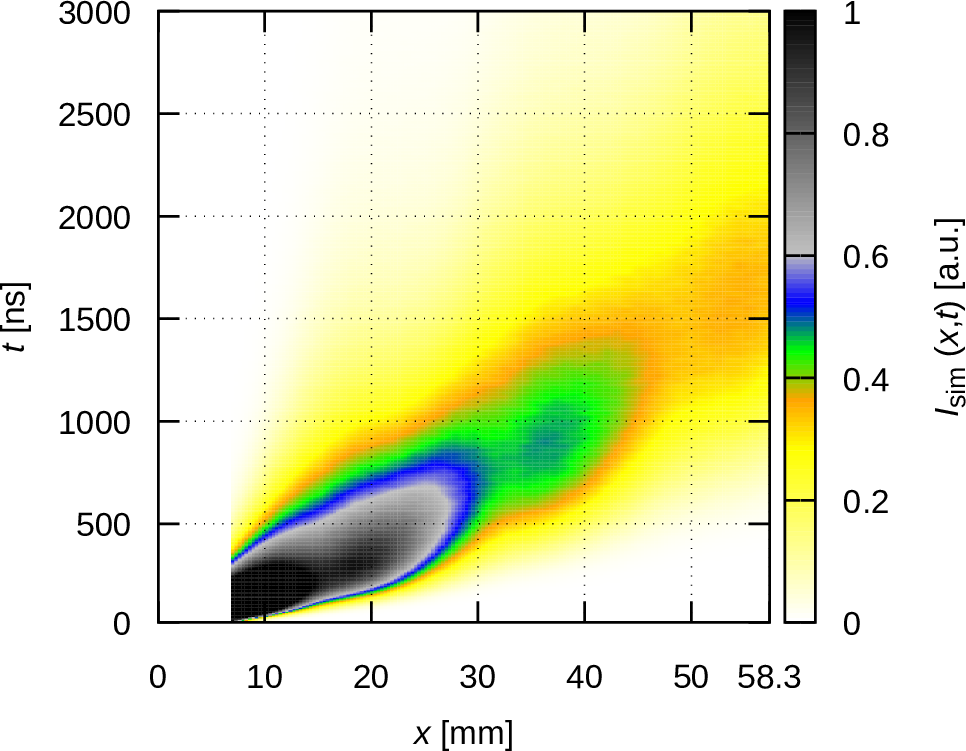}\\
\footnotesize{(b)}\includegraphics[width=0.45\textwidth]{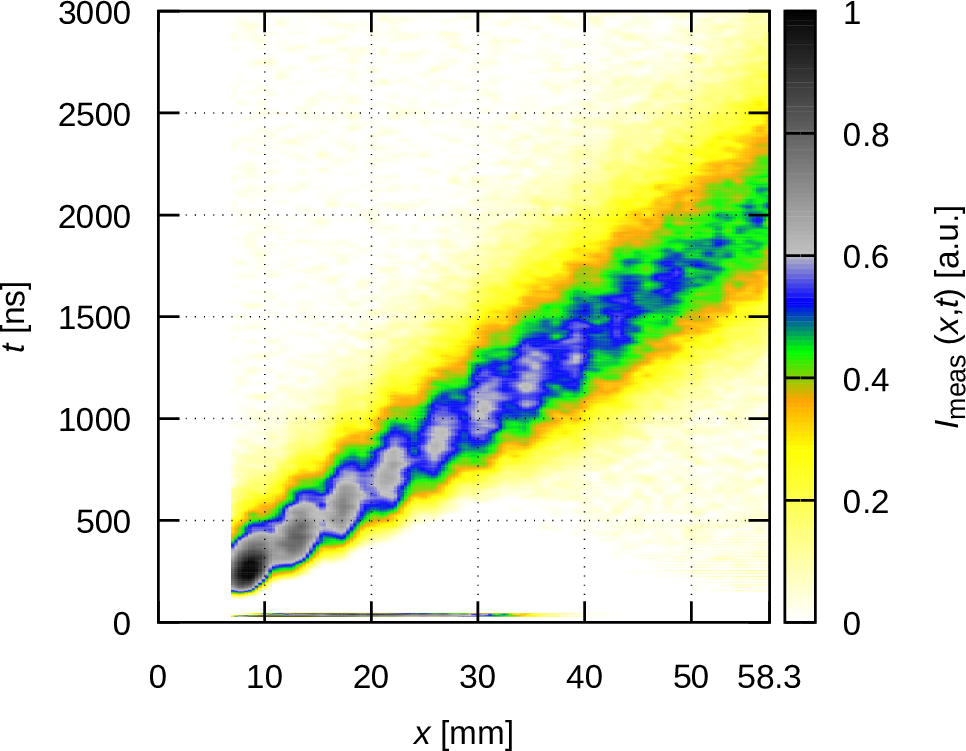}~
\footnotesize{(d)}\includegraphics[width=0.45\textwidth]{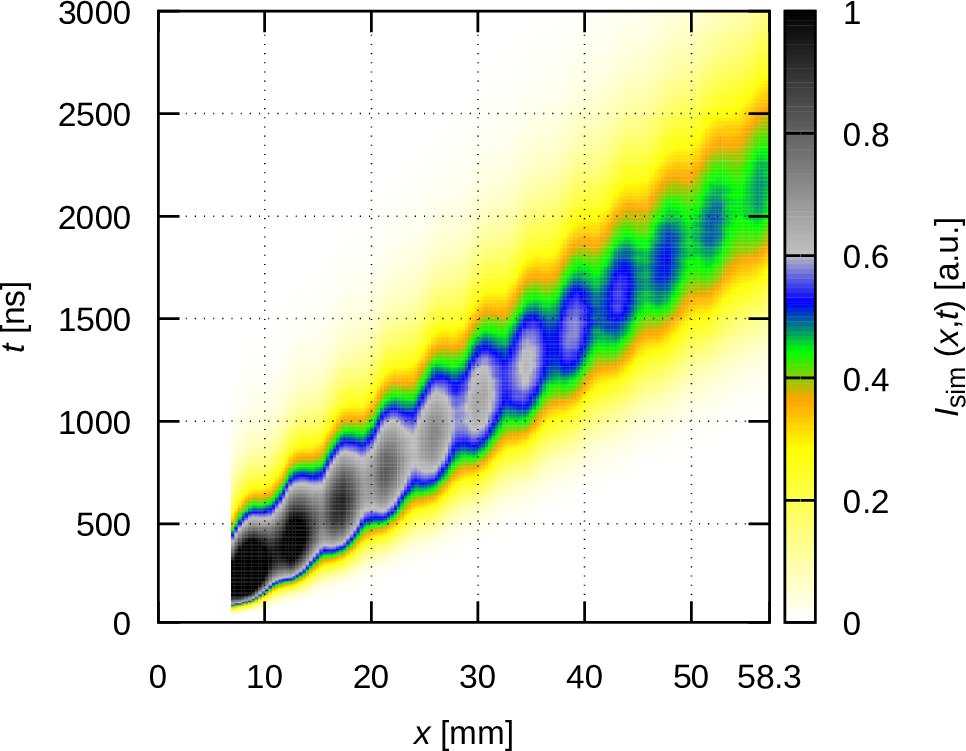}\\
\caption{Experimentally recorded detector current (a-b), at pressure values of 100 Pa and 400 Pa, respectively, and (c-d) corresponding simulation results, at $E/N$ = 30 Td.}
\label{fig:res2}
\end{figure}

Figure \ref{fig:res2} presents the results of the variation of the pressure at fixed $E/N$ = 30 Td. As expected, when a lower pressure of 100 Pa is used in the measurements (and in the corresponding simulations) compared to the 200 Pa case, for which the results were presented in figure \ref{fig:res1}(a) and (d), the density maxima are separated by a higher distance. Oppositely, at $p$ = 400 Pa, the periodicity of the maxima becomes two times more dense, as compared to the 200 Pa case. 

The experimental results presented above provide a direct way to observe the equilibration of electron swarms at moderate $E/N$ values, which was mostly studied only theoretically so far. The good agreement with the corresponding simulation results confirms the correctness of the data.

\subsection{Characterisation of the detector}

\label{sec:detector}

Now we turn to the analysis of the electrons' motion in the detector to answer the question: "What property of the swarm is measured by the detector?" To answer this question we need to pay attention to the electron trajectories in the detector region ("Region 2" in figure \ref{fig:simple-setup}).

\begin{figure}[h!]
\begin{center}
\includegraphics[width=0.5\textwidth]{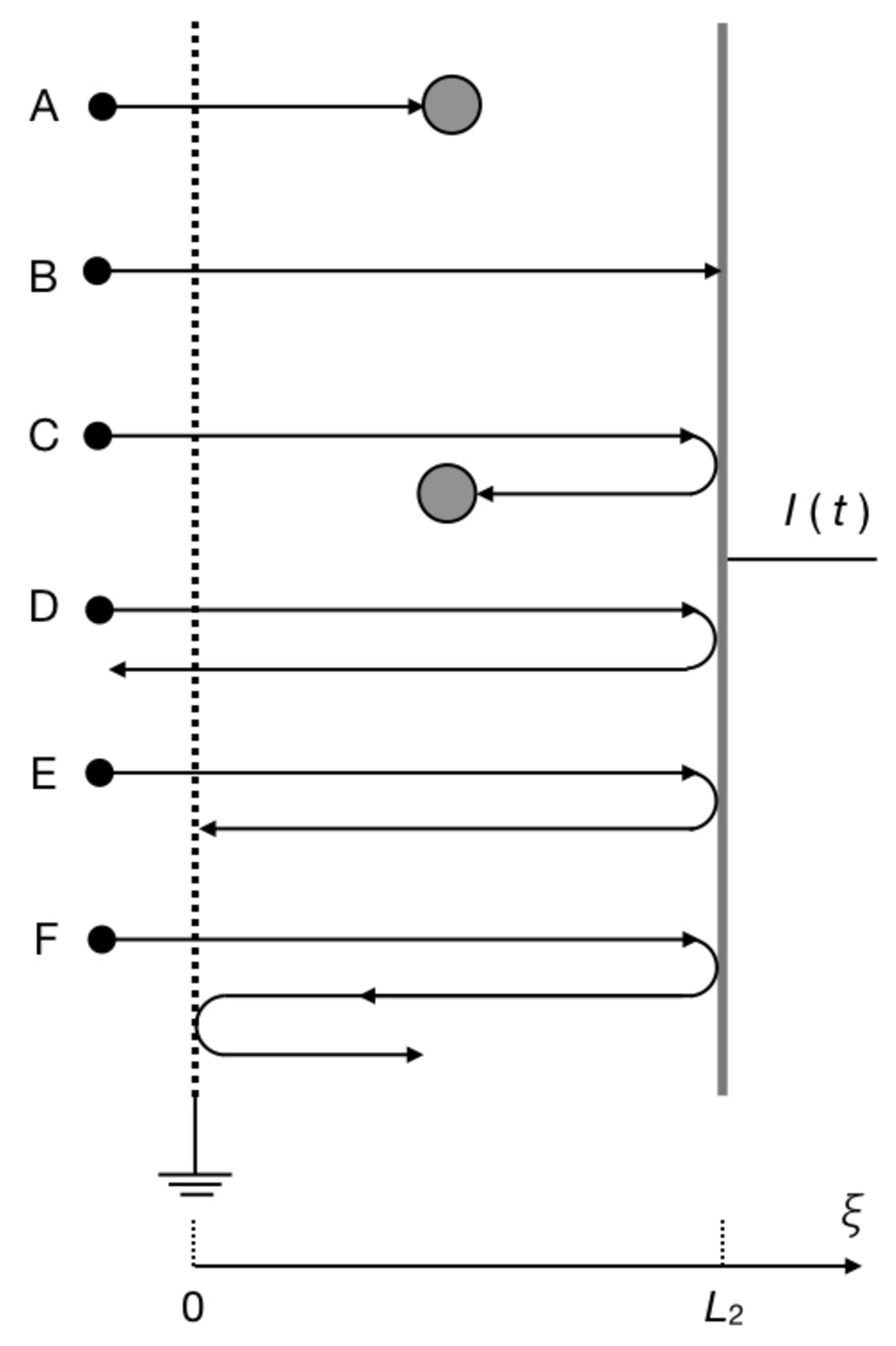}~
\end{center}
\caption{Basic types of electron trajectories (assuming that electrons enter the detector region have a velocity vector parallel to the $x$ direction. The small black circles represent incoming electrons, the big circles represent gas atoms with which the electrons collide.}
\label{fig:cases}
\end{figure}

Figure \ref{fig:cases} shows basic types of electron trajectories and their contributions to the detector current. In this analysis, we assume that (i) electrons pass through the mesh with a velocity vector that points in the $x$ direction,  (ii) all collision events results in isotropic scattering (as above), and (iii) electrons reaching the collector are reflected elastically and with a given probability (taken to be $P$=0.5, as above). Whenever we discuss a certain type of trajectory, we have in mind a large number of electrons (with similar energy and thus a similar collision free path length) that cross the mesh over some time interval that is (i) longer than the flight time of the electrons in the detector region, but (ii) much shorter than the period during which the whole electron cloud arrives at the detector.

If the collision free path is much shorter than the width of the detector region ($\lambda \ll L_2$) predominantly type A trajectories will occur. The free flight length of the electrons within the detector is in the order of $\lambda$. It is important to recognise that, as collisions result in isotropic scattering, these electrons give a detector current contribution only up to the first collision events, because after the collisions the direction of their velocities is randomised. Accordingly, the further transport of these electrons towards the two electrodes does not give a contribution to the measured current. 

With an increasing free path the electrons may reach the collector, where they may be absorbed (type B trajectories) or reflected (C--F-type trajectories). Note that reflected electrons give a negative contribution to the measured current as they move in the negative $x$ direction. The occurrence of type C trajectories is likely only when $\lambda \sim L_2$, as otherwise reflected electrons are again expected to have a long free path, that gives preference to the D-, E-, and F-type trajectories, which represent electron groups crossing the mesh (type D trajectories), being absorbed by the mesh (type E trajectories), or being reflected by the mesh (type F). Type F trajectories could be divided into further sub-types, however, this type of trajectory is not expected to occur frequently, as it requires reflection of electrons on the collector ($P = R$ = 0.5), interaction with the mesh ($P$ = 0.12, i.e. one minus the geometric transmission) and reflection there (assumed to have $P = R$ = 0.5), giving an overall probability of $P$ = 0.03. Thus F type trajectories may be excluded as major sources of the detector current. Returning to the D- and E-types of trajectories, these will give zero contribution on time average, for a large group of electrons.  

Thus in summary, 
\begin{itemize}
\item{whenever $\lambda \ll L_2$, the measured current will be proportional to the number of electrons entering the detector per unit time, i.e. their flux, the value of their $v_x$ velocity component (according to eq. (\ref{eq:ramo})) and to their flight time up to the first collision. As the product of the latter two is actually the path length, the current can be approximated as being proportional to $\lambda$;}
\item{whenever $\lambda \gg L_2$, the measured current will be proportional again to the flux of electrons entering the detector, the value of their $v_x$ velocity component and the probability of their absorption by the collector. Thus, a strongly reflecting collector will decrease the level of the measured signal for  electrons with a long free path.}
\end{itemize}

\begin{figure}[h!]
\footnotesize{(a)}\includegraphics[width=0.45\textwidth]{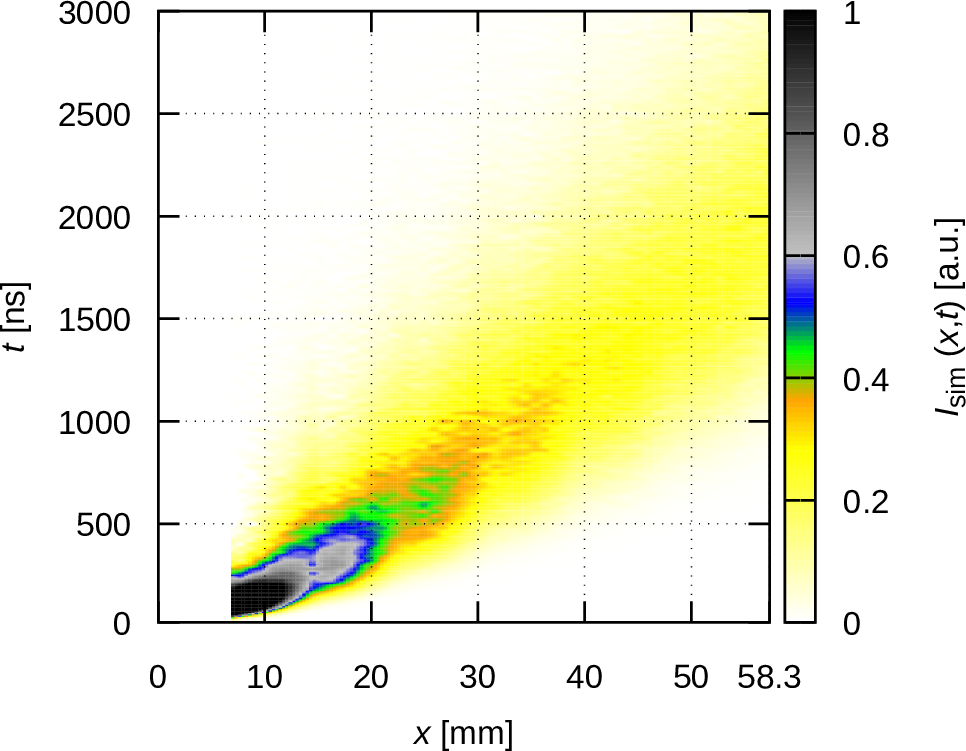}~
\footnotesize{(b)}\includegraphics[width=0.45\textwidth]{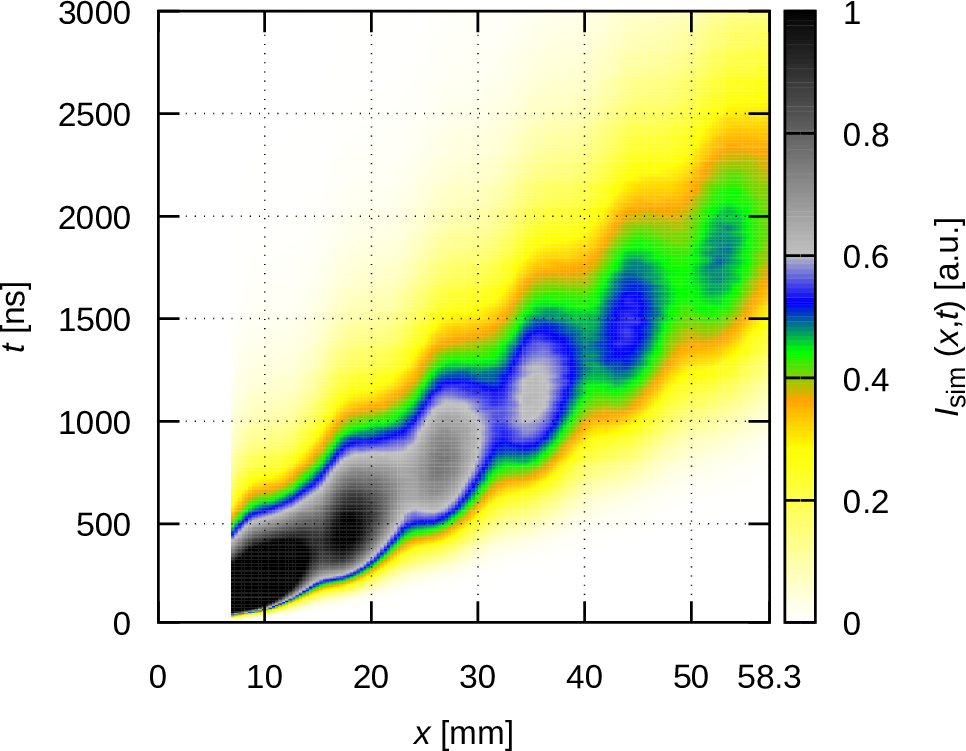}~
\caption{(a) Simulation results with (a) a highly reflecting ($R$ = 0.99) collector and (b) a low-reflection ($R$ = 0.01) collector, at $E/N$ = 30 Td and $p$ = 200 Pa.}
\label{fig:refl}
\end{figure}

This arguments are indeed confirmed by simulation results obtained with different values of electron reflectivity of the collector, presented in figure~\ref{fig:refl}. These data have been obtained at $E/N$ = 30 Td and 200 Pa argon pressure. The reflectivity values are $R$ = 0.99 for (a) and $R$ = 0.01 for (b). The conditions are the same in figure \ref{fig:res1}(b), for which $R$ = 0.5 was assumed. 

As it can be seen in figure~\ref{fig:refl}(a), the high reflectivity of the collector has a detriment effect on the detector signal, while a low reflectivity (figure~\ref{fig:refl}(b)) further increases the quality of the detector signal, beyond that shown in figure \ref{fig:res1}(b) for the realistic choice of $R$ = 0.5. These observations confirm the reasoning presented above and the good agreement between the experimental and simulation results shown in figure \ref{fig:res1} also confirms that the $R$ = 0.5 value, assumed for the reflectivity, is indeed realistic.

\begin{figure}[h!]
\begin{center}
\footnotesize{(a)}\includegraphics[width=0.45\textwidth]{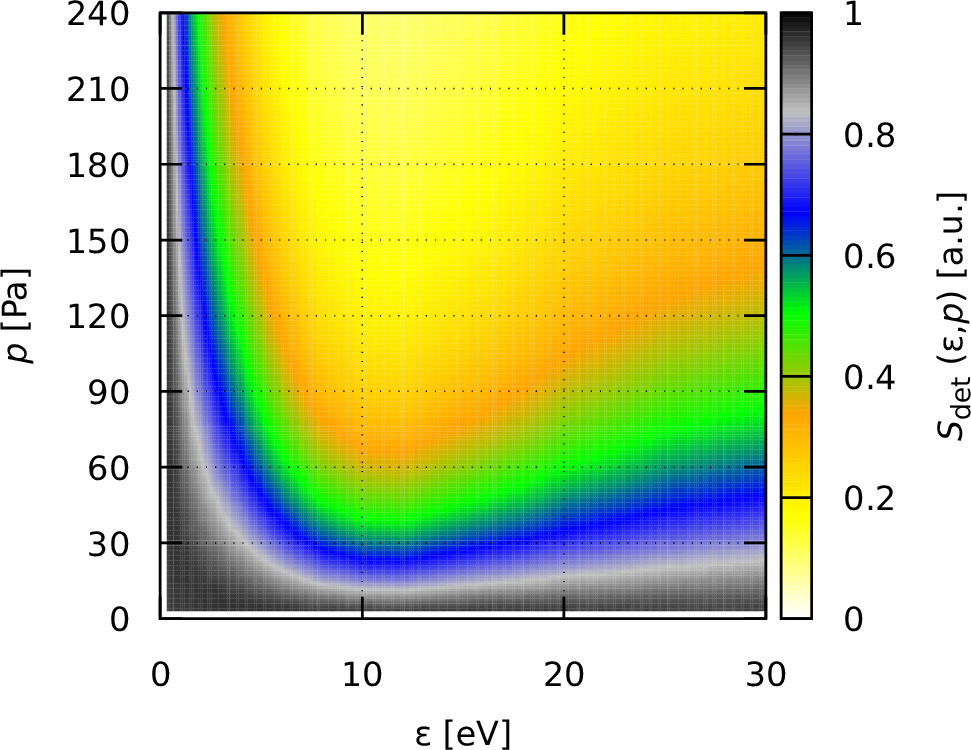}~~
\footnotesize{(b)}\includegraphics[width=0.45\textwidth]{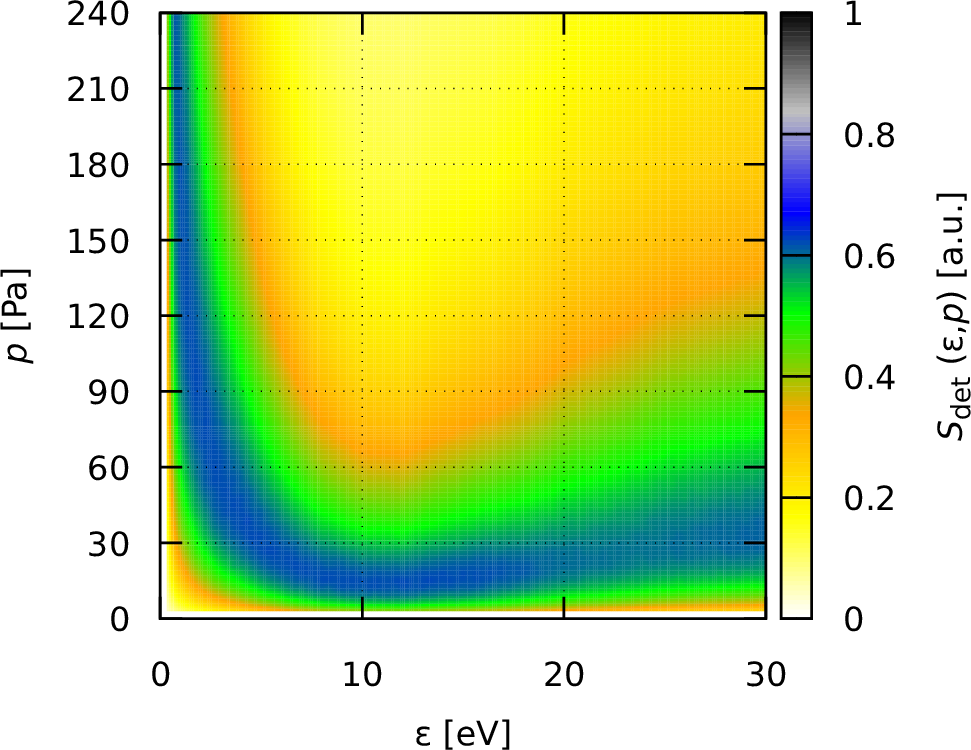}
\caption{Sensitivity of the detector, $S_{\rm det}$, (in arbitrary units) as a function of the energy of incoming electrons and the buffer gas pressure, for $R = 0.5$ (a) and for a highly reflecting collector with $R=0.99$ (b). An isotropic angular distribution of the incoming electrons is assumed.}
\label{fig:sens}
\end{center}
\end{figure}

We have conducted additional simulations to "measure" the sensitivity of the detector, $S_{\rm det}$. For this, a given number of electrons were injected into the detector region, at a defined pressure, with a defined energy, and isotropic distribution of initial velocity directions over the positive half sphere. This latter choice is justified by the fact the velocity distribution function of electrons is nearly isotropic at low to moderate $E/N$ values (as the average velocity is much smaller than the thermal velocity). The pressure and the energy were scanned over the domains 3 Pa -- 240 Pa and 0.4 eV -- 40 eV, respectively. For each pair of these parameters 10$^5$ initial electrons were used, the total current generated by these electrons served as a measure of the sensitivity of the detector. The results of these simulations are presented in figure~\ref{fig:sens}. 

Panel (a) shows the results for the case of a collector with $R=0.5$ reflection coefficient, a value that has been assumed in the simulations of the experimental system, while panel (b) shows the case of a highly reflecting collector, with $R=0.99$. We find that the sensitivity of the detector depends in a complicated manner on both the gas pressure and the energy of the incoming electrons. 

\begin{figure}[h!]
\footnotesize{(a)}\includegraphics[width=0.45\textwidth]{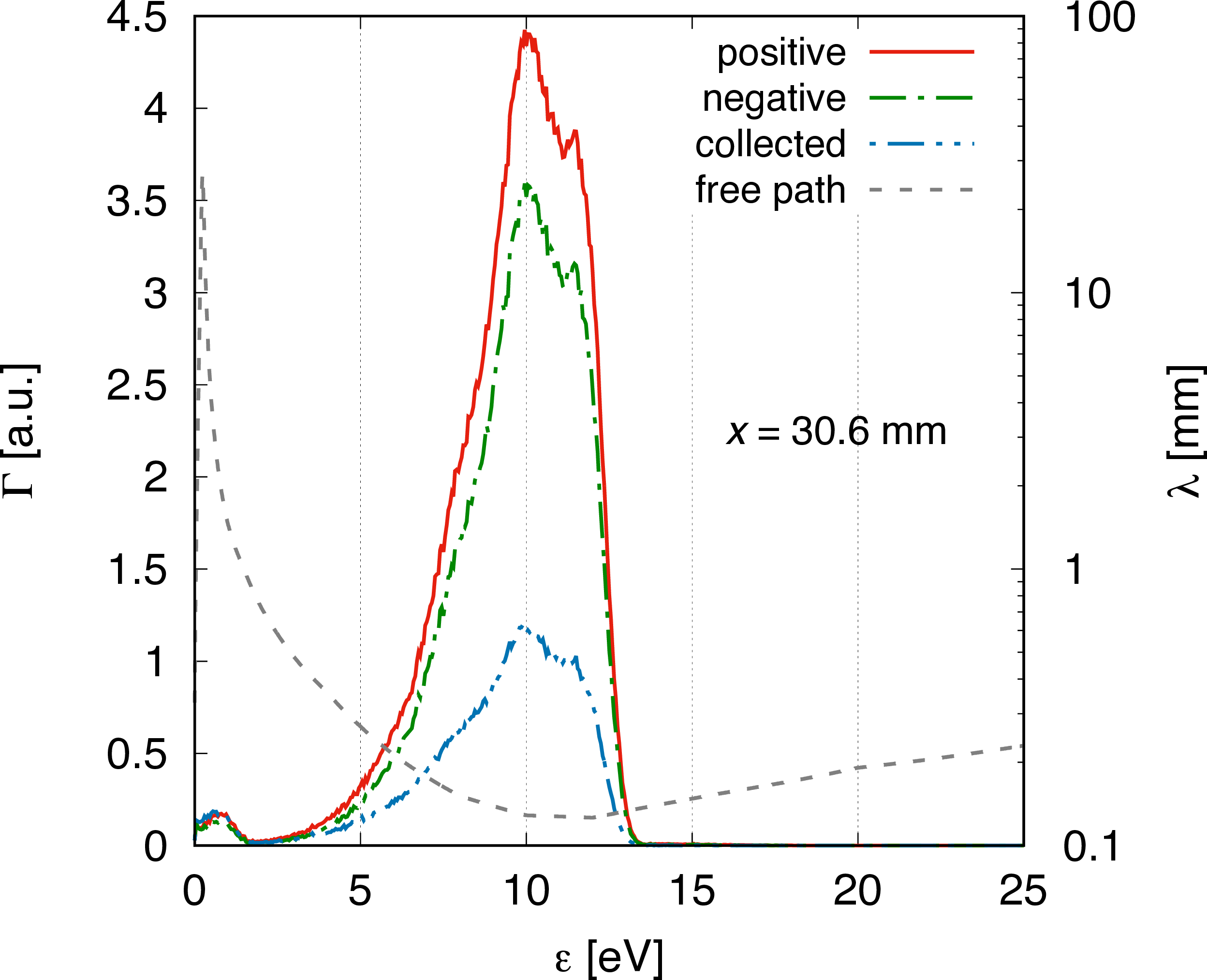}~
\footnotesize{(b)}\includegraphics[width=0.45\textwidth]{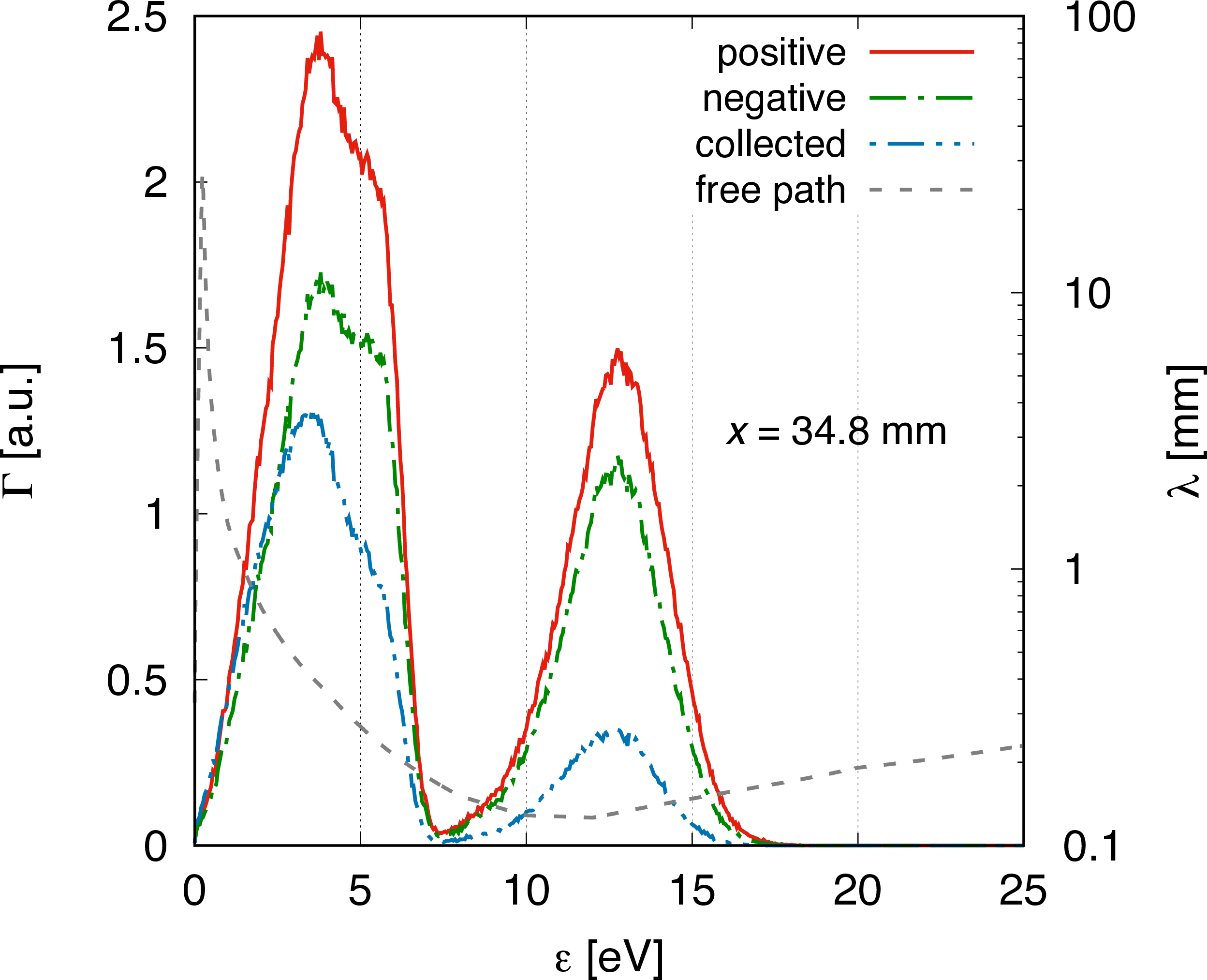}~
\caption{The distribution (flux) of the electrons according to their energy crossing the mesh in the positive and negative directions (labelled as "positive" and "negative", respectively), and absorbed at the collector (labelled as "collected"). $E/N$ = 30 Td, $p$ = 200 Pa, $R$ = 0.5. (a) $x$ = 30.6 mm and (b) $x$ = 34.8 mm. The free path ($\lambda$) of the electrons at the given gas pressure is shown on the right axis.}
\label{fig:flux}
\end{figure}

For the $R=0.5$ case we find a high sensitivity at low pressures. Up to about 5 Pa, the response of the detector is strong for all electron energies. At these low pressures, the majority of the electrons reaches the collector. Half of these electrons are absorbed, i.e. their trajectories are of type B (see figure~\ref{fig:cases}). The other half of the electrons will have trajectories of types C--F, which may decrease the response by $\sim$ 50\%. With an increasing pressure the detector sensitivity decreases, except for the electrons with very low energies. The electrons with energies $\sim$ 1 eV, or lower, still have a long free path (due to the Ramsauer minimum in the momentum transfer cross section) and many of them reach the collector and the above arguments apply to the types of their trajectories. For electrons with higher energies, however, the sensitivity drops and shows a minimum around 12 eV, where, actually the mean free path is the shortest (see later, in figure~\ref{fig:flux}). This drop of sensitivity is attributed to the A-type trajectories, which have gradually lower contributions to the detector current when the electron free flight becomes shorter at higher pressures. 

The reflectivity of the collector plays a central role in the sensitivity as the comparison of the panels of figure~\ref{fig:sens} reveals. The differences are concentrated, however, to the low-pressure domain, as at higher pressures, as discussed above, A-type trajectories form and the electrons do not reach the collector without collisions. Therefore, above $\approx$ 50 Pa, the panels of figure~\ref{fig:sens} look identical. At low pressures, however, the sensitivity of the detector decreases drastically (by about a factor of 10 at the lowest pressures covered) when the reflectivity of the collector is increased to 0.99. This is due to the fact that the B-type trajectories will be replaced by mostly D--F type trajectories, which result in a cancellation of the current created by the electrons moving into opposite directions.

We note that the results shown here are specific for argon gas, for any other gases the results for $S_{\rm det}$ may differ significantly because of the different cross sections. The existence of the Ramsauer minimum for argon, e.g., plays an important role in the behaviour of slow electrons in the detector region and influences the sensitivity considerably. 

The dependence of the sensitivity of the detector on the gas pressure and electron energy contributes to the slight shift of the structures seen in the density of the swarm in a plane-parallel configuration (figure \ref{fig:tof1}(b)) vs. the measured current in the experimental system and its computed counterpart (figures \ref{fig:res1}(a) and (d)). Figure \ref{fig:flux} displays, at two different positions, the energy resolved fluxes of electrons (i) entering Region 2 via the mesh (labelled as "positive"), (ii) being absorbed by the collector (labelled as "collected") and (iii) leaving the detector (Region 2) via the mesh, in the negative direction (labelled as "negative"). Panel (a) corresponds to a position, $L_1$ = 30.6 mm, where the measured current is minimum (see figure \ref{fig:res1}(a)), while panel (b) corresponds to a position, $L_1$ = 34.8 mm, where the current is maximum. In both cases $E/N$ = 30 Td, $p$ = 200 Pa, and the reflectivity of the detector is $R$ = 0.5. The mean free path of the electrons is also shown in figure \ref{fig:flux} (as curves labelled "free path"). At $L_1$ = 30.6 mm one major electron group enters the detector with energies between 5 eV and 13 eV. In this case (figure \ref{fig:flux}(a)) the free path is much smaller than $L_2$ = 1 mm, therefore only about a quarter of these electrons, is collected. Compared to this, at $L_1$ = 34.8 mm, two electron groups reach the detector. While for the high energy group the collection efficiency is also small, for the low energy group, as the free path is longer (see figure \ref{fig:flux}(b)) about the half of the electrons are collected. This shows a drastic change of the sensitivity of the detector, $S_{\rm det}$, as a function of spatial position under the actual conditions of the experiment, where swarm equilibration is studied.

\section{Summary}

\label{sec:sum}

We have investigated the equilibration of electron swarms in argon gas. Following the illustration of the general behaviour of electron swarm equilibration via numerical simulations of steady-state (SST) and time-dependent systems, we presented experimental investigations of the equilibration phenomenon by using a scanning drift tube apparatus that allows observation of the spatio-temporal development of electron swarms. The experimental studies have been complemented with numerical simulations of the experimental system. A very good agreement has been found between the measured and computed detector currents. 

We have also presented a detailed study of the operation of the detector by analysing types of possible electron trajectories and by carrying out simulations for the detector sensitivity as a function of electron energy and the gas pressures. This analysis has indicated a strong variation of the sensitivity on these two parameters, which explains the slight differences between the spatio-temporal distributions of electron density in the swarm and that of the measured detector current. These differences, thus, do not originate from uncertainties in the measurements and/or in the computations, but have well-defined reasons. Our studies provided an insight into the equilibration effects from the experimental side, complementing a number of previous theoretical / simulation studies.

\ack This work was supported by National Office for Research, Development and Innovation (NKFIH) via grants 119357 and 115805, by the DFG via SFB 1316 (project A4). SD and DB are supported by the Grants No. OI171037 and III41011 from the Ministry of Education, Science and Technological Development of the Republic of Serbia.


\section*{References}
\begin{thebibliography}{99}

\bibitem{DT1} Crompton R W, Elford M T and Jory R L 1967 {\it Aust. J. Phys.} {\bf 20} 369

\bibitem{DT2} Crompton R W 1972 {\it Aust. J. Phys.} {\bf 25} 409

\bibitem{DT3} Nakamura Y 1987 {\it J. Phys. D: Appl. Phys.} {\bf 20} 933

\bibitem{DT4} De Urquijo J, Arriaga C A, Cisneros C and Alvarez I 1999 {\it J. Phys. D: Appl. Phys.} {\bf 32} 41

\bibitem{DT5} Dahl D A, Teich T H and Franck C M 2012 {\it J. Phys. D: Appl. Phys.} {\bf 45} 485201

\bibitem{swarm1} Petrovi\'c Z L, Dujko S, Mari\'c D, Malovi\'c G, Nikitovi\'c \v{Z}, \v{S}a\v{s}i\'c O, Jovanovi\'c J, Stojanovi\'c V, Radmilovi\'c-Radenovi\'c M 2009 {\it J. Phys. D: Appl. Phys.} {\bf 42} 194002  

\bibitem{cross1} Tagashira H 1992 {\it Aust.J.Phys.} {\bf 45} 365

\bibitem{cross2} Petrovi\'c Z Lj, \v{Suvakov} M, Nikitovi\'c \v{Z}, Dujko S, \v{S}a\v{s}i\'c O, Jovanovi\'c J, Malovi\'c G and Stojanovi\'c V 2007 {\it Plasma Sources Sci. Technol} {\bf 16} S1

\bibitem{cross3} Morgan W L 1991 {\it Phys. Rev. A} {\bf 44} 1677

\bibitem{equi} Malovi\'c G, Strini\'c A, \v{Z}ivanov A, Mari\'c D, Petrovi\'c Z 2003 {\it Plasma Sources Sci. Technol.} {\bf 12} S1

\bibitem{Donko2011} Donk\'o Z 2011 {\it Plasma Sources Sci. Technol.} {\bf 20} 024001

\bibitem{per1} Kolobov V I and Arslanbekov R R 2006 {\it IEEE Trans. Plasma. Sci.} {\bf 34} 895 

\bibitem{per2} Dujko S, White R D and Petrovi\'c Z 2008  {\it J. Phys. D: Appl. Phys.} {\bf 41} 245205 

\bibitem{per3} White R D, Robson R E, Dujko S, Nicoletopoulos, P and Li B 2009 {\it J. Phys. D: Appl. Phys.} {\bf 42} 194001 

\bibitem{per4} Nicoletopoulos P and Robson R E 2008 {\it Phys. Rev. Lett.} {\bf 100} 124502

\bibitem{LiWR2002} Li B, White R D and Robson R E 2002 {\it J. Phys. D: Appl. Phys.} {\bf 35} 2914 

\bibitem{DujkoWPR2011} Dujko S, White R D, Petrovi\'{c} Z Lj and Robson R E 2011 {\it Plasma Source Sci. Technol.} {\bf 20} 024013

\bibitem{LiRW2006} Li B, Robson R E and White R D 2006 {\it Phys. Rev. E} {\bf 74} 026405

\bibitem{co2} Vass M, Korolov I, Loffhagen D, Pinh\~{a}o N, Donk\'o 2017 {\it Plasma Sources Sci. Technol.} {\bf 26} 065007 

\bibitem{rsi} Korolov I, Vass M, Bastykova N Kh, Donk\'o Z 2016 {\it Rev. Sci. Instrum.} {\bf 87} 063102

\bibitem{SH} Sirkis M D, Holonyak N (Jr) 1966 {\it American J. Physics} {\bf 34} 943

\bibitem{Hayashi} Hayashi M, {\it Recommended values of transport cross sections for elastic 
collision and total collision cross section for electrons in atomic and molecular gases} 
Report IPPJ-AM-19, Nagoya Institute of Technology (unpublished).

\bibitem{Robson} Robson R E 1991 {\it Aust. J. Phys.} {\bf 44} 685

\bibitem{SakaiTS1977} Sakai Y, Tagashira H and Sakamoto S 1977 {\it J. Phys. D: Appl. Phys.} {\bf 10} 1035

\bibitem{StojanovicP1998} Stojanovi\'{c} V D and Petrovi\'{c} Z Lj 1998 {\it J. Phys. D: Appl. Phys.} {\bf 31} 834

\end{thebibliography}
\end{document}